\ifx\documentclass\undefined
\documentstyle{elsart}
\newcommand\lesssim{\stackrel{\lower.7ex\hbox{$<$}}{\lower.7ex\hbox{$\sim$}}}
\newcommand\gtrsim{\stackrel{\lower.7ex\hbox{$>$}}{\lower.7ex\hbox{$\sim$}}}
\else
\documentclass{elsart}
\usepackage{amssymb}
\let\tenrm\relax
\let\cal\mathcal
\let\rm\mathrm
\fi
\begin{document}

\begin{flushright}
hep-ph/9503353\\
INLO-PUB-95/04\\
March 1995,\\
revised February 1996
\end{flushright}

\vskip2cm

\begin{frontmatter}
\title{Hard Photons in $W$ Pair Production at LEP 2}

\author{Geert Jan van Oldenborgh}
\thanks{Research supported by the Stichting FOM}
\address{{\tt gj@rulgm0.leidenuniv.nl}\\
  Instituut-Lorentz, Rijksuniversiteit Leiden,
  Postbus 9506, NL-2300 RA Leiden,
  Netherlands}

\begin{abstract}
The properties of hard photon radiation in $W$ pair production at LEP 2 are 
studied, with emphasis on the energy loss relevant to the $W$ mass 
measurement.  We use a combination of the exact \mbox{1-photon} matrix element 
and leading logarithmic structure functions.  Defining unobservable, 
observable and initial state photons in the phase space,  it is shown that 
neither the one-photon matrix element nor the leading logarithmic structure 
functions alone give an adequate description of the energy loss due to 
observable or initial state photons.  An event generator based on these 
calculations is available.
\end{abstract}
\journal{Nuclear Physics B}
\date{October 1995}
\end{frontmatter}
\clearpage


\section{Introduction}

One of the major goals of the LEP 2 experiments is an accurate 
determination of the mass of the $W$ boson.  A comparison of the 
measured value with the one predicted from other precision electro-weak 
measurements will serve as a sensitive  check of the Standard Model.
The most precise measurement method seems to be a constrained fit of the 
final state in the semi-leptonic (lepton+jets) and hadronic (4 jets) 
channels.  These fits assume that there is no hard radiation or that the 
photon spectrum is known.  Final state radiation can easily be 
incorporated by adding the photon to the nearest outgoing charged 
particle, but the shift in the fitted $W$ mass caused by initial state 
radiation will have to be determined  using theoretical input.  In 
particular, in order to get a precision in $M_W$ better than 10 MeV, the 
average energy loss has to be known to 20--25 MeV.

In $W$ pair production the distinction between initial and final state 
radiation is not unique (unlike in the case of neutral particles such 
as the $Z$ boson).   In the matrix element, the universal leading 
logarithmic parts (proportional to $\log(s/m_e^2)$) are easily 
separable, but the non-universal finite terms do not split naturally 
(although the current splitting technique \cite{Gentle} seems to achieve 
a `good' separation).
We therefore define initial state radiation in the phase space rather 
than the matrix element; this is also more in line with the 
experimental possibilities.
We use two definitions.  The first is the criterion that the outgoing 
photon is closer to the beam than to any charged final state particle. 
The second assumes that events with observable photons are analysed 
separately, and defines initial state radiation as unobservable 
photons closer to the beam than to other charges particles; this is 
almost equivalent to a cone around the beam.

The first computation of hard radiation to off-shell $W$ pair 
production was performed by Aeppli \& Wyler \cite{Andre&DanielBrems}; 
this calculation did not include mass effects so it could not cover 
the collinear regions.  In Ref.\ \cite{WWF} we gave the extension to 
the full phase space and an event generator.  Other event generators 
contain a leading log description of initial state radiation, like 
\cite{ExcaliburIni,WOPPER,jetset94}, where the last two use a 
shower algorithm to generate $p_T$ and multiple photons.  The 
semi-analytical calculation of Ref.\ \cite{Gentle} also gives some of 
the finite terms in the current splitting scheme, which allows for a 
gauge-invariant separation of initial and final state separation.  They 
include the universal non-resonant graphs.  Finally, a first exploration of 
the hard radiation off the single $W$ (t-channel) graphs of $e^+e^- \to e^- 
\bar{\nu}_e u \bar{d} \gamma$ was presented in Ref.\ \cite{KEKenudg}.
The goal of this article is to give a very good description of hard 
radiation, including the exact $1\gamma$ matrix element at large 
angles, and the leading ${\cal O}(\alpha^2)$ contributions at small 
angles to the beam.  The various definitions of initial state 
radiation are compared, and  the validity of the approximations made 
by other computations investigated.  The influence of the universal 
non-resonant diagrams is confirmed to be small; other diagrams leading 
to some of the 4-fermion + $\gamma$ final state, such as $ZZ$ graphs,  
single $W$ production and QCD graphs, have not yet been included.

We start with a technical description of the matrix elements and phase 
space used.  Next we give results for the energy and $p_T$ spectrum of 
observable, initial state and unobservable radiation, and the average 
energy loss.  This is compared with the same quantities in the one-photon 
and leading log approximation, and the influence of the universal 
non-resonant graphs is computed.  Finally, we mention where the event 
generator which incorporates these calculations can be obtained.


\section{Method}
\label{sec:method}

\subsection{Hard Bremsstrahlung}

The method used in this calculation is an exact matrix element 
for the reaction $e^+e^- \to 4f + \gamma$ 
convoluted with structure functions to describe multiple collinear 
photons.  The phase space is thus the one-photon phase space (the 
multi-channel MC described in Ref.\ \cite{WWF}) plus collinear 
photons, the sum of which is represented by one particle in each 
direction.  No attempt has been made to use a shower algorithm to 
resolve these into individual photons.

We use two matrix elements.  The first one is the original massless one of 
Ref.\ \cite{Andre&DanielBrems} (explicit expressions for this 
matrix element are also given in Ref.\ \cite{AndreThesis}).  Leading mass 
effects are added as described in Ref.\ \cite{WWF}:  When the angle between 
the photon and a charged particle $i$ is so small that the mass effects become 
significant we use
\begin{equation}
    |M_{\rm{massive}}|^2 = 
        |M_{\rm{massless}}|^2 \frac{(q_i q_\gamma)}{(p_i p_\gamma)}
        - e^2 \frac{m_i^2}{(p_i p_\gamma)^2}
\end{equation}
where the $q_j$ denote the massless momenta, and the $p_j$ the massive ones.  
This way the logarithmically mass singular and finite ($m_i^2/m_i^2$) terms 
are included correctly.

The second matrix element, used mainly for cross-checks, 
has been generated automatically using MadGraph \cite{MadGraph}.  
It is an order of magnitude slower but contains all mass effects without 
approximations.  
Using this matrix element one also has the possibility to study the 
effect of universal non-resonant graphs.  The $ZZ$ and single $W$ graphs are 
also included, but as long as the appropriate channels have not been 
added to the phase space mappings these can not be used.\footnote{A 
possible exception is the study of large-angle electrons in the case of single 
$W$ production, as in Ref.\ \cite{KEKenudg}.}

Given the number of expected events these $1\gamma$ matrix elements 
are accurate enough for a good description of large-angle hard photon 
emission.  However, they fall short in describing the collinear initial 
state bremsstrahlung, which is dominated by large logarithms 
$\log(s/m_e^2)$.
These can be resummed (and higher order contributions can be included) in 
structure functions (transition functions) \cite{LeidenEemumu2loop,%
KuraevFadinStruct,AltarelliMartinelliStruct,TrentadueStruct,KleissIni}, but 
only after integrating out the transverse momentum $p_T$.  As in Ref.\ 
\cite{EEWW}, we combine these two approaches by only resumming the 
radiation inside a cone 
$\theta < \theta_c$; subtracting the leading log ${\cal O}(\alpha)$ part 
from the explicit matrix element to avoid double counting.  
This gives multi-photon 
emission, but only on the side where the explicit photon happened to be.

Using structure functions one would like to write the matrix element as
\begin{eqnarray}
    \sigma & = & \int dk^+ \rho(k^+) \int dk^- \rho(k^-) 
        \frac{|M\bigl((1-k^+)(1-k^-)s\bigr)|^2}{2(1-k^+)(1-k^-)s}
\;,
\end{eqnarray}
with $k^\pm$ the fraction of the incoming momentum taken away by 
collinear initial state radiation off the $e^\pm$ and $M(\hat{s})$ some 
hard scattering amplitude.  We use the structure functions given in Refs 
\cite{LeidenEemumu2loop,Jochem&Fred&KarolW}\footnote{The 
difference with the 3-loop YFS structure function \cite{SJ91} is 
${\cal O}(10^{-3})$ in the variables studied.}
\begin{equation}
\label{eq:rho}
    \rho(k) = \beta k^{\beta-1}\Bigl(1 + \delta_1^{\rm{V+S}} + 
        \delta_2^{\rm{V+S}} \Bigl) + \delta_1^{\rm{H}}(k) + 
        \delta_2^{\rm{H}}(k)
\;,
\end{equation}
with $\beta = \frac{\alpha}{\pi}\bigl(L-1\bigr)$ and $L = 
\log(\mu^2/m_e^2)$ the large collinear logarithm.  The terms $\delta$ 
denote the infrared finite parts of the leading one- and two-loop 
corrections:
\begin{eqnarray}
    \delta_1^{\rm{V+S}}  & = & \frac{\alpha}{2\pi}
        \bigl(\frac{3}{2}L + \frac{\pi^2}{3} - 2\bigr) \\
    \delta_1^{\rm{H}}(k) & = & \frac{\alpha}{2\pi}(1-L)(2-k) \\
    \delta_2^{\rm{V+S}}  & = & \bigl(\frac{\alpha}{2\pi}\bigr)^2
        \Bigl(\frac{9}{8} - \frac{\pi^2}{3}\Bigr)L^2 \\
    \delta_2^{\rm{H}}(k) & = & \bigl(\frac{\alpha}{2\pi}\bigr)^2
        \Bigl[ -\frac{1+(1-k)^2}{k}\log(1-k)
\nonumber\\&&\qquad\mbox{}
        + (2-k)\bigl(\frac{1}{2}\log(1-k) - 2\log(k)
        - \frac{3}{2}\bigr) - k\Bigr]L^2
\end{eqnarray}
For the scale we use $\mu^2 = s(1-\cos\theta_c)/2$, which reproduces the 
${\cal O}(\alpha)$ results and agrees with the scale usually taken when no 
cone is defined.

We have to define the hard scattering amplitude in such a way that 
collinear photon radiation is not included twice, preferably keeping it 
positive definite.  The procedure followed resembles the one described 
in Ref.\ \cite{EEWW}.  For large angles (larger than 
$\theta_c$) we take for this amplitude the $1\gamma$ matrix 
element.  However, inside the cone we must first subtract from this the 
${\cal O}(\alpha)$ leading log part which has already been included in 
the structure functions:
\begin{equation}
    |M_{\rm{sub}}|^2 = 
        e^2 \left(\frac{1+(1-k)^2}{k(1-k)}\frac{1}{(p_i p_\gamma)} 
            - \frac{m_i^2}{(p_i p_\gamma)^2} \right) 
        |M_0\bigl((1-k)s\bigr)|^2
\;.
\end{equation}
Here $k = E_\gamma/E_e$ and $|M_0|^2$ denotes the matrix element 
without photon evaluated at a lower energy.  This expression only holds 
strictly in the collinear limit, and breaks down when the cone is chosen too 
large (typically around $90^\circ$).

Next we add again the exponentiated leading log result: the lowest order 
matrix element times structure function, but now at $k' = k^\pm + x$, 
with $x = E_\gamma/E_e$ the fraction of the beam energy taken by the 
explicit photon and the direction $(\pm)$ determined by the direction of 
the explicit photon:
\begin{eqnarray}
    \lefteqn{|M_{\rm{add}}\bigl((1-k^+)(1-k^-)s\bigr)|^2
         \stackrel{\theta_\gamma<\theta_c}{=} }
\nonumber\\
    && S \int_{E_{\rm{min}}/E_e} \!\!\!\!\!\! dx \;
        \frac{|M^{(0)}\bigl((1-k^\mp)(1-k')\bigr)|^2}{1-k'}\,
        \frac{\rho(k')}{\rho(k^\pm)} \,
        \frac{g(x,k')}{\int_{E_{\rm{min}}/E_e}^{k'} dx' g(x',k')}
\;.
\end{eqnarray}
$g(x,k')$ is an arbitrary function and $S$ a symmetry factor defined later.  
In order to get roughly the same behaviour as the original matrix element we 
choose
\begin{equation}
    g(x,k') \approx 
        \frac{1}{1-k'} \rho(k'-x) \, \rho_2\bigl(x/(1-k'+x)\bigr)
\;,
\end{equation}
with $\rho_2(x)$ the function $\rho(x, \alpha\to2\alpha)$ which 
approximates the matrix element squared in one direction.  We use an 
approximation as the function has to be analytically integrable.

However, this is only half the story, as the lower bound on the photon 
energy is now always imposed on the side on which the explicit photon 
happened to be.
The region in phase space where the collinear photon on the other side 
has satisfied this bound, $k^\mp E_e > E_{\rm{min}}$, is included by 
defining $S = 2-\theta(E_{\rm{min}} - k^\mp E_e)$.  The 
step function avoids double counting when both collinear photons have 
large energies.
The total matrix element is then given by
\begin{equation}
    |M_{\rm{full}}|^2 = |M_{\rm{massive}}|^2 - |M_{\rm{sub}}|^2 + 
        |M_{\rm{add}}|^2
\end{equation}

The collinear energy lost in the forward and backward directions is the 
sum of many photons.  We can approximate the $p_T$ of this bunch of 
collinear photons by the $p_T$ spectrum of a single photon of the same 
energy to obtain a smooth transition from the $1\gamma$ matrix element 
to the exponentiated one.  The resulting violations of energy-momentum 
conservation are of order $E_\gamma \sin^2\theta$, and hence negligible 
for small cone sizes.  The implementation of a true parton shower algorithm 
\cite{WOPPER} is foreseen for the near future.

The same subtraction of the leading logarithmic contribution can be 
performed for final state photons.  As the exact difference between
a single fermion and a fermion accompanied by collinear photons is not 
important for most measurements, we have not performed the exponentiation as 
we did for the initial state photons.  Also, radiation off a hadronic final 
state is badly described by the radiation of an on-shell quark: the effective 
mass of the radiating particle is the mass of the whole jet, not the on-shell 
mass of a naked quark.  This reduces the bremsstrahlung considerably.  In the 
time domain this means that the quark will radiate off gluons before it can 
radiate off a photon; the amount of hard radiation near a jet is thus 
overestimated.  The possibility to turn off the final state radiation means 
that one can leave the generation of these photons to more heuristic 
algorithms (see, e.g., \cite{jetset94}) tuned to the observed spectrum if this 
spectrum is important.

An unsolved problem is the matter of gauge invariance of 
brems\-strahlung off un\-stable particles.  As we use a constant width 
in our calculations the current is conserved (the substitution 
$p^\mu_\gamma \to \epsilon^\mu_\gamma$ gives zero), but this is not 
sufficient to guarantee gauge invariance. However, there are no 
cancellations which could amplify the gauge violating terms, so we 
expect the effects to be small.  A test using the inclusion of triangle 
graphs is being performed.

\subsection{Soft Bremsstrahlung}

In the following we will consider mainly the average energy lost by photons, 
\begin{equation}
\label{eq:Eave}
    \langle E_\gamma \rangle = \frac{1}{\sigma_0^{\rm{isr}}} 
        \int \! dE_\gamma \; E_\gamma \; \frac{d\sigma}{dE_\gamma}
\end{equation}
The normalisation $\sigma_0^{\rm{isr}}$ is computed from the tree level matrix 
element for the reaction $e^+e^- \to 4$ fermions, convoluted with structure 
functions.  This has an accuracy of a few percent.  However, as an overall 
factor it will not influence the conclusions very much.

In the case of a strict one-photon calculation the integral in Eq.\ 
\ref{eq:Eave} is not sensitive to soft photons, and we can make accurate 
predictions without knowing the virtual and soft corrections.  However, when 
we add the possibility of multiple photon emission through the procedure given 
above, and define $E_\gamma$ as the {\em sum} of the photon energies, these do 
play a role.  Note that we consider the regime $E_{\rm{min}} \ll 
\Gamma_W$.  In this case the soft corrections are easily found to be
\begin{equation}
    |M_{\rm{soft}}|^2 = e^2 \sum_{i,j} Q_i Q_j B(p_i,p_j;E_{\rm{min}},\lambda) 
        |M_0|^2
\;,
\end{equation}
with $Q_i$ the charges of the particles in terms of the unit charge $e$.  The 
function
\begin{equation}
    B(p_i,p_j;E_{\rm{min}},\lambda) = \pm \frac{(p_i p_j)}{(2\pi)^3} 
        \int_{E_\gamma < E_{\rm{min}}} \frac{d^3p_\gamma}{2E_\gamma} 
        \frac{1}{(p_i p_\gamma)} \frac{1}{(p_j p_\gamma)}
\end{equation}
is given in Ref.\ \cite{tHooft&Veltman} (the sign depends on whether the 
particles are incoming or outgoing).  It depends logarithmically on 
the infra-red regulator, for which we use a small photon mass $\lambda$.
We can include the corresponding infra-red logarithms of the virtual 
corrections by the substitution $\lambda \to \mu_{\rm{IR}}$.  We choose for 
this scale $\mu_{\rm{IR}} = M_W/n$, with $n$ the number of charged final state 
particles.  With this heuristic, the difference in the total cross section 
between the full result and the leading logarithmic approximation is less than 
1.5\% in the region $175 < \sqrt{s} < 205$.

\begin{figure}
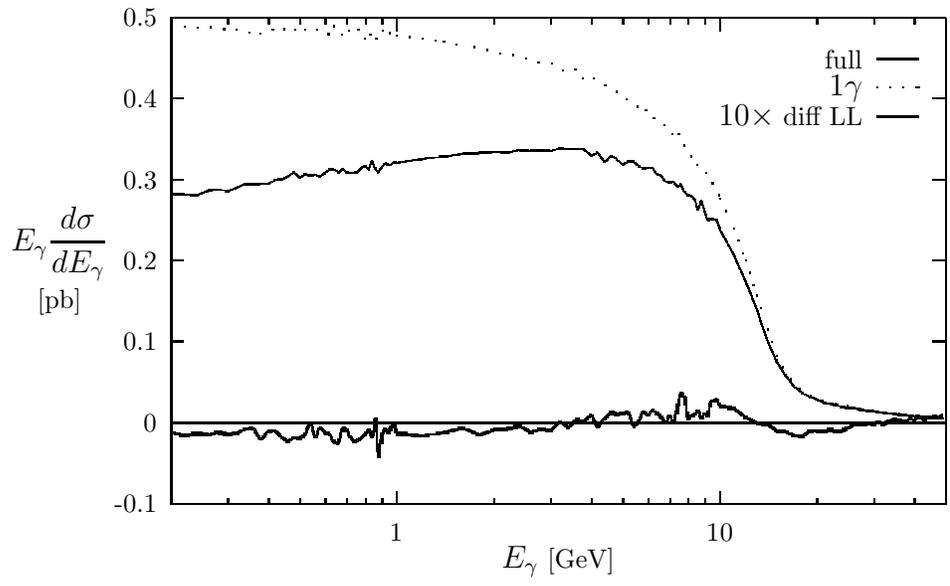

\begin{center}
\setlength{\unitlength}{0.240900pt}
\ifx\plotpoint\undefined\newsavebox{\plotpoint}\fi
\sbox{\plotpoint}{\rule[-0.200pt]{0.400pt}{0.400pt}}%
%
\end{center}
\caption{Initial state photon energy in the semi-leptonic channel.}
\label{fig:isr}
\end{figure}

Both the soft and virtual corrections will contain logarithms in the 
virtuality of the $W$ bosons, $\log(p^2-M_W^2+iM_W\Gamma_W)$.  However, these 
cancel in the sum.  This can be seen in the hard photon spectrum 
(Figs~\ref{fig:obs}--\ref{fig:isr}) as an absence of a kink at $E_\gamma 
\approx \Gamma_W$.  The cut-off on the initial-final and final-final state 
interference as the energy increase is thus exactly compensated by the 
emergence of radiation off the $W$ bosons.

We also extract the Sudakov double logs
\begin{equation}
    |M_{\rm{sud}}|^2 = \frac{e^2}{(2\pi)^2} \sum_{i} Q_i^2 
        \log^2 \frac{m_i}{\mu_{\rm{IR}}} \; |M_0|^2
\;.
\end{equation}
When we introduce the structure functions we have to subtract again the part 
which would be double-counted:
\begin{equation}
    |M_{\rm{sub,soft}}|^2 = \frac{e^2}{2\pi^2}
        \bigl[2(L-1)\log \frac{2E_{\rm{min}}}{\sqrt{s}} + \frac{3}{2}L +  
            \frac{\pi^2}{3} - 2\bigr]
\end{equation}
The tree term has already been added to the hard radiation for $E_\gamma > 
E_{\rm{min}}$, we need to add it only to the soft terms as
\begin{equation}
    |M_{\rm{tree}}|^2 = 
        \bigl(1 - \theta((k_{\rm{min}} - k^+)(k_{\rm{min}} - k^-)\bigr) 
        |M_0|^2
\end{equation}
Adding all up the estimate for the non-radiative part is
\begin{equation}
    |M_{\rm{full,soft}}|^2 = |M_{\rm{soft}}|^2 -
        |M_{\rm{sud}}|^2 - |M_{\rm{sub,soft}}|^2 + |M_{\rm{tree}}|^2
\end{equation}
The other terms in the virtual corrections do not contain large logarithms, 
and are thus expected to be of order $\alpha/\pi \lesssim 1\%$.  As they only 
enter as corrections to the main effects of hard radiation in the quantities 
studied in this paper this is negligible.
For the factorizable graphs, for which the off-shell result is under study, 
this reproduces the full result to a few percent in the cross section.
A change in the scale $\mu_{\rm{IR}}$ of a factor 2 gives a difference of 
$3\%$ in the cross section, but only $1\%$ in the observable energy (where the 
effect of the virtual and soft terms is much smaller).  We verified that the 
numbers obtained with this soft matrix eleemnt did not any more depend on the 
cut-off $E_{\rm{min}}$.


\section{Results}

We now turn to the predictions we can make for the hard photon spectrum 
in $W$ pair production.  We first define the `LEP 2 canonical cuts' that 
are used to compute observables that can easily be compared to other 
calculations.  These cuts also give an impression what an actual 
experiment could observe.  Next we show the energy and $p_T$ spectrum for 
observable and initial state photons.  This can be compared with the 
one-photon approximation on the one hand, and the leading log 
approximation on the other.  Finally we discuss the effect of the 
non-resonant diagrams.

The canonical cuts of the ADLO/TH detector \cite{Canonical} are given in 
Table \ref{tab:cuts}.  $\tau$'s are not considered.  The cuts 
on the photon separations should be interpreted as follows: first each 
photon is combined with the nearest charged particle if the angle to it 
is smaller than $5^\circ$ (outgoing) or $1^\circ$ (beam); combining with 
the beam means dropping it.  Next all the other cuts are applied.  This 
way spurious collinear divergences near cuts are avoided.

\begin{table}[hbt]
\begin{center}
\begin{tabular}{|r|cccc|}
\hline
separation & $\gamma$  & $e$       & $\mu$     & jet \\
\hline
$\gamma$   & $5^\circ$ & $5^\circ$ & $5^\circ$ & $5^\circ$ \\
$e$        & $5^\circ$ & $5^\circ$ & $5^\circ$ & $5^\circ$ \\
$\mu$      & $5^\circ$ & $5^\circ$ & $5^\circ$ & $5^\circ$ \\
jet        & $5^\circ$ & $5^\circ$ & $5^\circ$ & 5 GeV \\
\hline
\end{tabular}
\begin{tabular}{|r|rr|}
\hline
acceptance & $E_{\rm{min}}$ & $\theta_{\rm{min}}$ \\
\hline
$\gamma$   & 0.1 GeV       & $ 1^\circ$ \\
$e$        &   1 GeV       & $10^\circ$ \\
$\mu$      &   1 GeV       & $10^\circ$ \\
jet        &   3 GeV       & $ 0^\circ$ \\
\hline
\end{tabular}
\end{center}
\caption{Canonical cuts}
\label{tab:cuts}
\end{table}

Observable photons are defined as those which pass these cuts.  Unobservable 
initial state photons are the ones cut out by the energy cut or the minimum 
angle to the beam; the sum of their energies and $p_T$ can be deduced from 
momentum conservation.  Finally, we define initial state photons to be the 
unobservable ones plus observable photons which are closer to the beam than to 
any other charged particles.

In the subsequent part of this work we use the following parameters. We 
work in the $G_\mu$ scheme, but with the coupling of the extra photon 
determined by $\alpha(\mu^2=0)$.  All masses are taken from the Particle 
Data Group 1994 edition \cite{PDG94}.  The quark masses were taken 
$m_u=63$ MeV, $m_d = 83$ MeV, $m_s = 215$ MeV and $m_c = 1.5$ GeV for 
compatibility with the virtual off-shell calculation.  The $W$ mass is 
80.23 GeV.  The $W$ width is computed in leading order, with the strong 
coupling constant $\alpha_s = 0.117$; this gives $\Gamma_W = 2.081$ GeV.  
To conserve the electromagnetic current we use a fixed $W$ width; the pole 
position is computed as $\mu_W = M_W^2 - \Gamma_W^2 - 5/2\pi\,\Gamma_W^3/M_W
-iM_W\Gamma_W - i\Gamma_W^2/pi + i(1-1/pi^2)\Gamma_W^3/M_W$.
The $Z$ width is neglected, and the 
energy of the outgoing photon restricted to less than $E - (M_Z + 
5\Gamma_Z)^2/4E$ to avoid the $Z$ peak.\footnote{The cross section above 
this cut is entirely negligible \cite{Andre&DanielBrems}.}  Unless 
otherwise noted we assumed a beam energy of 87.5 GeV.  We used a cone angle of 
$10^\circ$.  Changing this to $5^\circ$ shifts the observables by an amount 
comparable to the integration accuracy, ${\cal O}(1\%)$.

\begin{figure}[b]
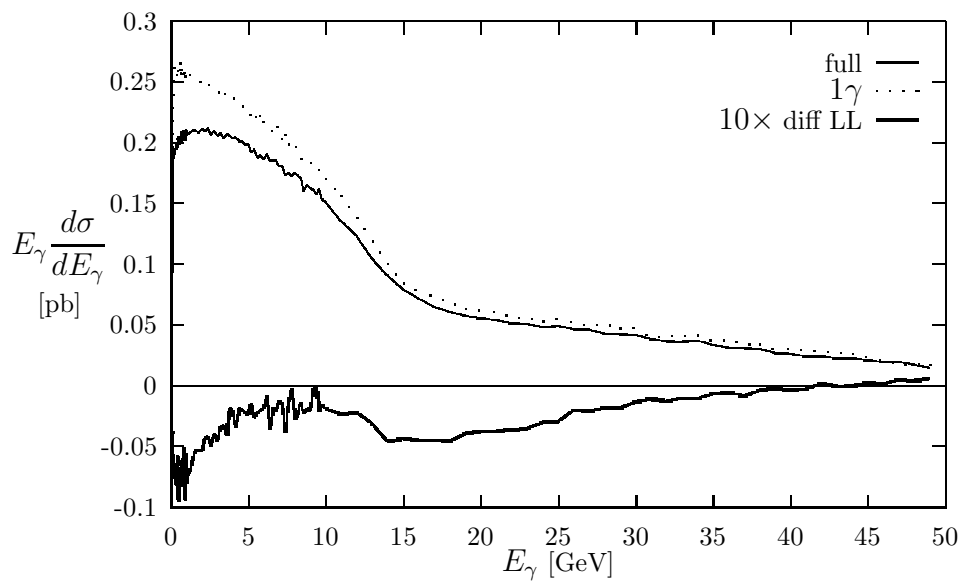

\begin{center}
\setlength{\unitlength}{0.240900pt}
\ifx\plotpoint\undefined\newsavebox{\plotpoint}\fi
\sbox{\plotpoint}{\rule[-0.200pt]{0.400pt}{0.400pt}}%

\end{center}
\caption{Observable photon energy in the semi-leptonic channel.}
\label{fig:obs}
\end{figure}

The results are presented in Figs \ref{fig:obs}--\ref{fig:isr} and 
Tables \ref{tab:175} and \ref{tab:205}.  First we give the observable 
tree level non-radiative cross section ($\sigma_0$), the same convoluted 
with leading log structure functions ($\sigma_0^{\rm{isr}}$) and with 
non-resonant graphs ($\sigma^{\rm{isr}}_{0+\rm{nr}}$).  For the next 
entries we consider the full calculation described in section 
\ref{sec:method}, the exact one-photon matrix element ($1\gamma$) and 
the leading log result (LL).  For technical reasons the leading 
logarithmic approximation can only be computed for photons emitted in 
the forward direction ($\theta_c=90^\circ$), so the scale is taken to be 
$\mu^2 = s/2$. This will underestimate the leading log result.  We also 
give the full result including the universal non-resonant diagrams 
(full$+$nonres).  The statistical errors on the cross sections are 
${\cal O}(0.1\%)$, but the differences $(\mbox{LL} - \mbox{full})$ and 
$((\mbox{res}+\mbox{nonres}) - \mbox{res})$ were computed directly in this 
form and have relative errors of a few percent and a few tens of percents 
respectively on these differences.  
The statistical errors on the the average photon 
energies are slightly larger, 0.3--0.5\%.  We introduced an upper cut-off on 
the photon energy to avoid the $Z$ peak, this influences only the observable 
average energy.

For the normalisation we need the total cross section for $W$ pair 
production.  As long as the full off-shell one-loop result is not yet 
available we use the lowest order cross section with leading logarithmic 
structure functions.  As the final state corrections cancel against the 
corrections to the width, and the initial-final interference is expected 
to be of order $\alpha\Gamma_W$ this is a reasonable approximation.  
This uncertainty only affects the overall normalisation of the results, 
not the differences between the different results.  The main effect of 
the initial state radiation is of course to lower the result, as we are 
just above the threshold for $W$ pair production and phase space is 
limited.  The contribution from the non-resonant graphs is negligible in 
this energy range (as had already been observed in Ref.\ 
\cite{Andre&DanielBrems}).  With the scale choice used, the full result 
deviates from this estimate by less than 1.5\%.

One can see from the Tables that an appreciable fraction (around one quarter) 
of the events will be accompanied by photons observable in the ADLO/TH set of 
cuts.  This is due to the excellent forward coverage ($\theta_\gamma > 
1^\circ$) and electromagnetic calorimeter ($E_\gamma>0.1$ GeV) assumed in the 
canonical cuts.   Reducing the angle to $10^\circ$ this fraction still is 
around 20\%, of which half also has an $E_\gamma > 1$ GeV.  For the $W$ mass 
measurement it would be advantageous to make use of the extra information 
present and analyse these events separately.  Neither the $1\gamma$ matrix 
element nor the leading log approximation give a satisfactory description of 
the observable photons.  By radiating only one photon one effectively 
normalises to the non-radiative lowest order, which is too large.  On the 
other hand, the leading log approximation misses the negative initial-final 
state interference terms ($E_\gamma \lesssim \Gamma_W$) and the radiation off 
the (off-shell) $W$ bosons (mainly when $E_\gamma \gtrsim 
\Gamma_W$).\footnote{The fact that there is no discernible structure at 
$E_\gamma\approx\Gamma_W$ in Fig.\ \ref{fig:obs} suggests that these two 
contributions are intimately connected; indeed, one cannot define the 
`radiation off the $W$' in a gauge invariant way.}  Finally, given that the 
un-exponentiated large-angle contribution of the cross section still is 20\% 
of the total cross section the region $E_\gamma < 1$ GeV should not be trusted 
to 1\% even in the full calculation.  Most of the cross section here is, 
however, associated with the final state and hence does not influence the $W$ 
mass measurement.

The observable photon energy is dominated by final state radiation and hence 
not very interesting.  The unobservable energy spectrum is much more 
independent of the final state.  It is not completely independent due to the 
possibility of observable jets in the beam pipe ($\theta_j=0^\circ$).  The 
initial state radiation associated with these events is sometimes associated 
with the final state by the canonical cuts, thus lowering the average energy.  
As the radiation off jets is not modelled correctly anyway, a cut will have to 
be imposed to exclude this contamination.  One sees that analysing observable 
photons separately reduces the average energy loss, and hence the size of the 
theoretical corrections to be applied to the fitted $W$ mass.  The difference 
in $\langle E_\gamma^{\rm{isr}}\rangle$ is due to the amount of phase space 
available: 2, 3 or 4 charged particles.
For the unobservable radiation the leading log approximation is, as expected, 
quite good\footnote{The systematic shift for leptonic, semi-leptonic and 
hadronic channels is probably due to the choice of the scale $\mu_{\rm{IR}}$, 
which should be slightly different for the different final states 
considered.}.  As in the total cross section, the contribution of the 
non-resonant graphs is not visible due to the limited integration
accuracy.  The contributions from the universal non-resonant graphs cannot be 
distinguished from the statistical fluctuations,\footnote{The numbers 
presented here are based on $2.5\;10^5$ weighted events per channel and took 
about 100 hours in all to compute on a fast workstation.} but may just be 
relevant.
At 205 GeV the integrated radiative corrections are smaller, 
but as there is much more phase space for the photons, more can be 
observed.  The same remarks on the accuracy of the various 
approximations hold here, except that the deviations tend to be larger.


\section{Conclusions}

We have studied the properties of hard radiation in $W$ pair production 
at LEP with a combination of the exact matrix element for one-photon 
emission and resummed structure functions for the initial state 
collinear large logarithms.  We find that a large fraction of the $W$ 
events will be accompanied by observable hard photons; a separate 
analysis of these events should decrease the systematic errors in the 
$W$ mass measurement.  These events are not well described either by the 
$1\gamma$ matrix element or a pure leading log approximation.  The 
treatment of photons near jets is an unsolved problem.

The average energy lost by the remaining, unobservable photons is smaller.  
The canonical LEP 2 cuts leave a contamination of final state radiation due to 
the inclusion of jets down to the beam pipe, which will have to be removed.  
In this region of phase space, a leading log treatment is adequate.  We have 
not yet addressed the issues of radiation off non-resonant graphs other than 
the (probably negligible) universal ones; in particular the QCD, $ZZ$ and 
single $W$ graphs have not yet been included.

An event generator based on the calculations presented in this article 
can be obtained from {\tt ftp://rulgm4.LeidenUniv.nl/pub/gj} or via 
the World Wide Web.

\ack
I would like to thank Niels J{\o}rgen Kjaer and Willy van Neerven for comments 
on the manuscript and useful discussions, and the NIKHEF in Amsterdam for the 
use of their computers.



\begin{table}
\begin{center}
{\large$E_{\rm{beam}} = 87.5$ GeV}\\
$\displaystyle
\begin{array}{|r@{\;\;}|@{\;\;}rlrlrl|}
\hline
& \multicolumn{2}{l}{\mbox{leptonic}}
& \multicolumn{2}{l}{\mbox{semi-leptonic}}
& \multicolumn{2}{l|}{\mbox{hadronic}}\\
\hline
\sigma_0 [{\rm{pb}}] &    .724&(20\%) &    4.564&(20\%) &    7.132&(20\%) \\
\sigma_0^{\rm{isr}} [{\rm{pb}}] &    .604&&    3.803&&    5.944& \\
\sigma_{0+\rm{nr}}^{\rm{isr}} [{\rm{pb}}] &    .605&( .13\%) &    
3.808&( .14\%) &    5.952&( .14\%) \\
\hline
\sigma^{\rm{obs}}/\sigma_0^{\rm{isr}}&&&&&&\\
\mbox{full} &    .280&&     .265&&     .242& \\
1\gamma &    .359&(28\%) &     .322&(21\%) &     .297&(23\%) \\
\mbox{LL} &    .292&( 4.2\%) &     .273&( 2.9\%) &     .245&( 1.2\%) \\
\mbox{with nonres} &    .281&( .17\%) &     .265&( .10\%) &     .242&( .08\%) \\
\hline
\langle E_\gamma^{\rm{obs}}\rangle \mbox{ [GeV]} &&&&&&\\
\mbox{full} &   1.247&&    1.013&&     .777& \\
1\gamma &   1.414&( .167) &    1.145&( .132) &     .884&( .107) \\
\mbox{LL} &   1.265&( .018) &    1.039&( .025) &     .814&( .037) \\
\mbox{with nonres} &   1.245&&    1.012&&     .776& \\
\hline
\langle E_\gamma^{\rm{unobs}}\rangle \mbox{ [GeV]} &&&&&&\\
\mbox{full} &    .666&&     .665&&     .647& \\
1\gamma &    .864&( .198) &     .856&( .191) &     .857&( .210) \\
\mbox{LL} &    .663&(-.003) &     .659&(-.007) &     .657&( .010) \\
\mbox{with nonres} &    .665&&     .665&&     .646& \\
\hline
\langle E_\gamma^{\rm{isr}}\rangle \mbox{ [GeV]} &&&&&&\\
\mbox{full} &   1.223&&    1.174&&    1.063& \\
1\gamma &   1.488&( .264) &    1.419&( .245) &    1.320&( .257) \\
\mbox{LL} &   1.208&(-.015) &    1.176&( .002) &    1.088&( .025) \\
\hline
\end{array}$\end{center}
\caption[]{Cross section for observable 
photons and energy lost to observable, 
unobservable and initial state photons at 
$\sqrt{s} = 175$ GeV.}
\label{tab:175}
\end{table}
\begin{table}
\begin{center}
{\large$E_{\rm{beam}} = 102.5$ GeV}\\
$\displaystyle
\begin{array}{|r@{\;\;}|@{\;\;}rlrlrl|}
\hline
& \multicolumn{2}{l}{\mbox{leptonic}}
& \multicolumn{2}{l}{\mbox{semi-leptonic}}
& \multicolumn{2}{l|}{\mbox{hadronic}}\\
\hline
\sigma_0 [{\rm{pb}}] &    .831&( 7\%) &    5.280&( 8\%) &    8.316&( 8\%) \\
\sigma_0^{\rm{isr}} [{\rm{pb}}] &    .774&&    4.908&&    7.724& \\
\sigma_{0+\rm{nr}}^{\rm{isr}} [{\rm{pb}}] &    .774&(.07\%) &    
   4.912&(.08\%) &    7.730&(.08\%) \\
\hline
\sigma^{\rm{obs}}/\sigma_0^{\rm{isr}}&&&&&&\\
\mbox{full} &    .295&&     .289&&     .267& \\
1\gamma &    .354&(20\%) &     .326&(13\%) &     .302&(13\%) \\
\mbox{LL} &    .317&( 7.2\%) &     .303&( 4.9\%) &     .276&( 3.3\%) \\
\mbox{with nonres} &    .296&( .08\%) &     .289&( .08\%) &     .268&( .02\%) \\
\hline
\langle E_\gamma^{\rm{obs}}\rangle \mbox{ [GeV]} &&&&&&\\
\mbox{full} &   2.010&&    1.802&&    1.555& \\
1\gamma &   2.106&( .096) &    1.830&( .028) &    1.593&( .038) \\
\mbox{LL} &   2.084&( .074) &    1.853&( .051) &    1.599&( .044) \\
\mbox{with nonres} &   2.008&&    1.801&&    1.554& \\
\hline
\langle E_\gamma^{\rm{unobs}}\rangle \mbox{ [GeV]} &&&&&&\\
\mbox{full} &   1.902&&    1.923&&    1.884& \\
1\gamma &   2.266&( .364) &    2.240&( .316) &    2.241&( .357) \\
\mbox{LL} &   1.866&(-.035) &    1.880&(-.044) &    1.875&(-.009) \\
\mbox{with nonres} &   1.900&&    1.922&&    1.882& \\
\hline
\langle E_\gamma^{\rm{isr}}\rangle \mbox{ [GeV]} &&&&&&\\
\mbox{full} &   3.168&&    3.147&&    2.976& \\
1\gamma &   3.549&( .380) &    3.447&( .300) &    3.344&( .368) \\
\mbox{LL} &   3.156&(-.013) &    3.106&(-.041) &    2.992&( .017) \\
\hline
\end{array}$\end{center}
\caption[]{Cross section for observable 
photons and energy lost to observable, 
unobservable and initial state photons at 
$\sqrt{s} = 205$ GeV.}
\label{tab:205}
\end{table}

\clearpage
\section*{Figures}

Inclusion of the universal non-resonant graphs gives curves which are 
indistinguishable from the `full' calculations.  The wiggles are due to 
the limited statistical accuracy ($10^7$ weighted events per curve). 

\begin{figure}
\begin{center}
\setlength{\unitlength}{0.240900pt}
\ifx\plotpoint\undefined\newsavebox{\plotpoint}\fi
\sbox{\plotpoint}{\rule[-0.200pt]{0.400pt}{0.400pt}}%
%
\end{center}
\caption{Unobservable photon energy in the semi-leptonic channel.}
\label{fig:unobs}
\end{figure}


\end{document}